\documentclass{iaus}
\newcommand\kms{{\rm\,km\,s^{-1}}}
\newcommand\msun{\rm\,M_\odot}

\usepackage{graphicx}

\title[High-velocity runaway stars from three-body encounters] 
{High-velocity runaway stars from three-body encounters}

\author[V.V.Gvaramadze, A.Gualandris \and S.Portegies Zwart]   
{V.V.Gvaramadze$^1$, A.Gualandris$^2$ \and S.Portegies Zwart$^3$}

\affiliation{$^1$Sternberg Astronomical Institute, Moscow State University,
       Universitetskij Pr. 13, Moscow 119992, Russia \\ email: {\tt vgvaram@mx.iki.rssi.ru}
\\ [\affilskip] $^2$Center for Computational Relativity and Gravitation,
       Rochester Institute of Technology, 85 Lomb Memorial Drive, Rochester NY 14623, USA\\email:
       {\tt alessiag@astro.rit.edu} \\ [\affilskip] $^3$ Astronomical Institute `Anton Pannekoek' and Section
       Computational Science, Amsterdam University, Kruislaan 403, 1098 SJ Amsterdam, the Netherlands \\
       email: {\tt spz@science.uva.nl}}

\pubyear{2009}
\volume{266}  
\pagerange{1--6}
\jname{Star Clusters -- Basic Galactic Building Blocks throughout Time and Space}
\editors{R. de Grijs \& J. Lepine, eds.}
\begin{document}

\maketitle

\begin{abstract}
We performed numerical simulations of dynamical encounters between
hard massive binaries and a very massive star (VMS; formed through
runaway mergers of ordinary stars in the dense core of a young
massive star cluster), in order to explore the hypothesis that this
dynamical process could be responsible for the origin of
high-velocity ($\geq 200-400 \, \kms$) early or late B-type stars.
We estimated the typical velocities produced in encounters between
very tight massive binaries and VMSs (of mass of $\geq 200 \,
\msun$) and found that about $3-4$ per cent of all encounters
produce velocities of $\geq 400 \, \kms$, while in about 2 per cent
of encounters the escapers attain velocities exceeding the Milky
Ways's escape velocity. We therefore argue that the origin of
high-velocity ($\geq 200-400 \, \kms$) runaway stars and at least
some so-called hypervelocity stars could be associated with
dynamical encounters between the tightest massive binaries and VMSs
formed in the cores of star clusters. We also simulated dynamical
encounters between tight massive binaries and single ordinary
$50-100\, \msun$ stars. We found that from 1 to $\simeq 4$ per cent
of these encounters can produce runaway stars with velocities of
$\geq 300-400 \, \kms$ (typical of the bound population of
high-velocity halo B-type stars) and occasionally (in less than 1
per cent of encounters) produce hypervelocity ($\geq 700 \, \kms$)
late B-type escapers. 
\end{abstract}

\firstsection 

\section{Introduction}

The origin of high-velocity runaway stars can be attributed to two
basic processes: (i) disruption of a tight massive binary following
the (asymmetric) supernova explosion of one of the binary components
(Blaauw \cite{bl61}) and (ii) dynamical three- or four-body
encounters in dense stellar systems (Poveda, Ruiz \& Allen
\cite{po67}). In the first process, the maximum velocity attained by
runaway stars depends on the magnitude of the kick imparted to the
stellar supernova remnant [either a neutron star (NS) or a black
hole (BH)] and for reasonable values of this magnitude, the runaway
velocity does not exceed $\sim 200 \, \kms$ (e.g. Portegies Zwart
\cite{po00}; cf. Gvaramadze \cite{gv09}). In the second process, the
ejection velocity could be higher. For example, the maximum velocity
that a runaway star can attain in binary-binary encounters is equal
to the escape velocity from the surface of the most massive star in
the binaries and could be as large as $\sim 1400 \, \kms$ (Leonard
\cite{le91}).

The recent discovery of the so-called hypervelocity stars (HVSs;
e.g. Brown et al. \cite{br05}) -- the ordinary stars moving with
velocities exceeding the Milky Way's escape velocity -- attracted
attention to dynamical processes involving the supermassive BH in
the Galactic Centre (e.g. Gualandris, Portegies Zwart \& Sipior
\cite{gu05}; Baumgardt, Gualandris \& Portegies Zwart \cite{ba06}).
These processes [originally proposed by Hills (\cite{hi88}) and Yu
\& Tremaine (\cite{yu03})] can result in ejection velocities of
several $1000\kms$. Similar processes but acting in the cores of
young massive star clusters (YMSCs) and involving dynamical
encounters with intermediate-mass ($\sim 100-1000 \, \msun$) BHs
(IMBHs) were considered by Gvaramadze, Gualandris \& Portegies Zwart
(\cite{gva08}) to explain the origin of extremely high-velocity
($\sim 1000 \, \kms$) NSs (e.g. Chatterjee et al. \cite{ch05}) and
HVSs. Gualandris \& Portegies Zwart (\cite{gu07}) proposed that
exchange encounters between hard binaries and an IMBH formed in the
core of a YMSC in the Large Magellanic Cloud could be responsible
for the origin of the HVS HE\,0437$-$5439. A strong support to the
possibility that at least some HVSs originate in star clusters
rather than in the Galactic Centre comes from the proper motion
measurements for the HVS HD\,271791, which constrain the birth place
of this star to the outer parts of the Galactic disc (Heber et al.
\cite{he08}).

\section{High-velocity runaway stars from three-body encounters}

In this paper, we explore the hypothesis (Gvaramadze \cite{gv07})
that some high-velocity runaway stars could attain their peculiar
velocities in the course of strong dynamical encounters between hard
massive binaries and a very massive ($\geq 100-150 \, \msun$) star
(VMS), formed in the core of a YMSC through collisions and mergers
of ordinary massive stars (e.g. Portegies Zwart \& McMillan
\cite{po02}). In this process, one of the binary components is
replaced by the VMS, while the second one is ejected (the so-called
exchange encounter), sometimes with a high velocity. Our goal is to
check whether or not this process can produce early B-type stars
(the progenitors of NSs) with velocities of $\geq 200-400 \, \kms$
(typical of pulsars) and $3-4 \, \msun$ stars with velocities of
$\geq 300-400 \, \kms$ (typical of some late B-type halo stars).

In our study we proceed from the similarity between encounters
involving a VMS and those involving an IMBH (the latter process is
already known to be able to produce high-velocity runaways;
Gvaramadze et al. \cite{gva08}) and the fact that the radii of VMSs
are smaller than the tidal radii of the intruders (the massive
binaries), so that the tidal breakup and ejection can occur before
the binary components merge with the VMS (Gvaramadze \cite{gv07};
Gvaramadze et al. \cite{gva09}). Our study is motivated by the
recent evolutionary models of VMSs (Belkus et al. \cite{be07};
Yungelson et al. \cite{yun08}), which suggest that VMSs can lose
most of their mass via copious winds and leave behind IMBHs with
masses of $\leq 70 \, \msun$, which are not large enough to
contribute significantly to the production of high-velocity runaway
stars (see Gvaramadze et al. 2008). We therefore explore the
possibility that a VMS could produce high-velocity escapers (either
early or late B-type stars) before it finished its life in a
supernova and formed a BH. To check this possibility, we performed
numerical simulations of three-body exchange encounters using the
{\tt sigma3} package, which is part of the {\tt STARLAB} software
environment (McMillan \& Hut \cite{mc96}; Portegies Zwart et al.
\cite{po01}). For a detailed description of the simulations see
Gvaramadze et al. (\cite{gva09}).

\begin{figure}[h]
\begin{minipage}[h]{0.44\linewidth}
\center{\includegraphics[width=1\linewidth]{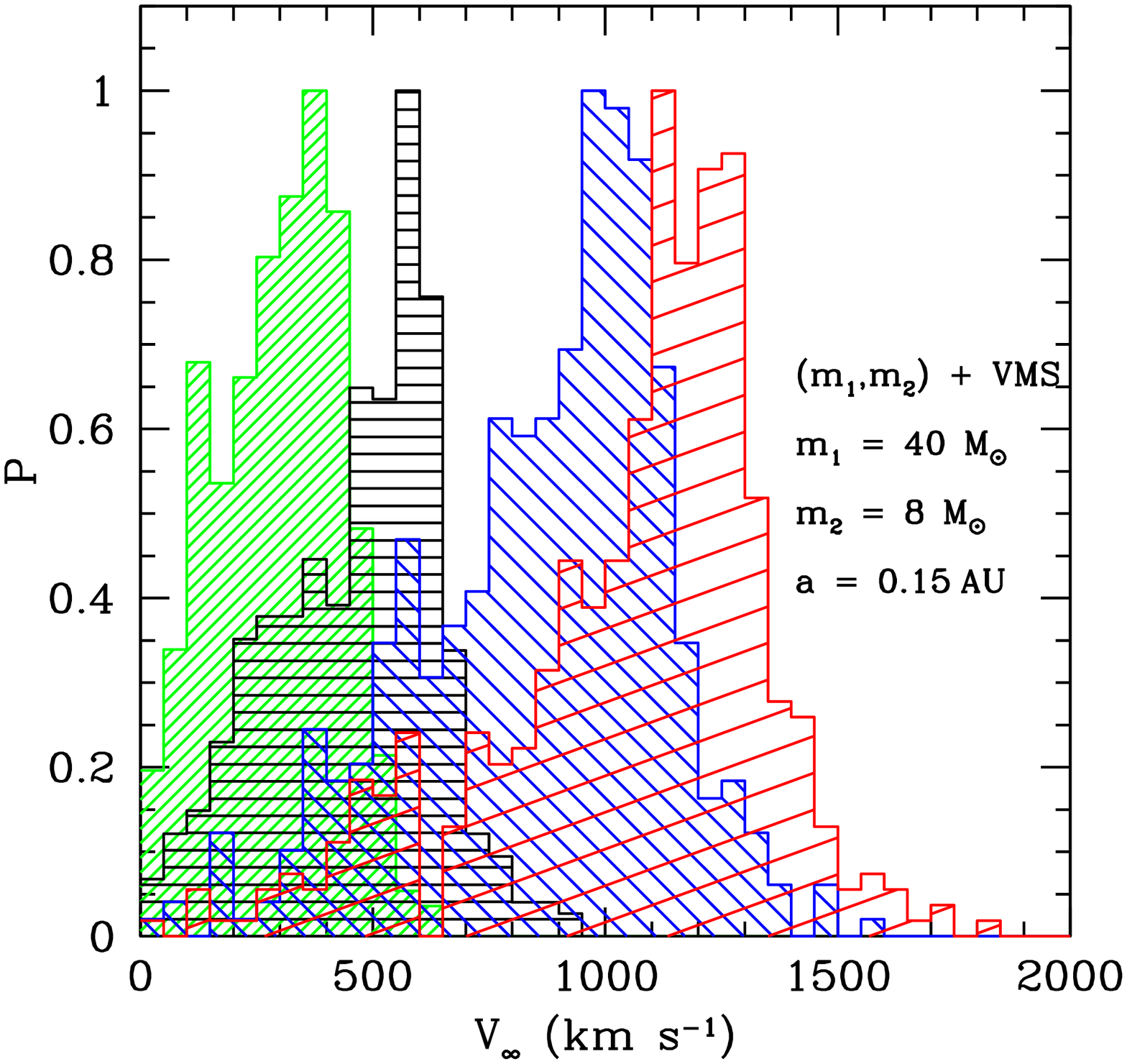}}
\end{minipage}
\hfill
\begin{minipage}[h]{0.44\linewidth}
\center{\includegraphics[width=1\linewidth]{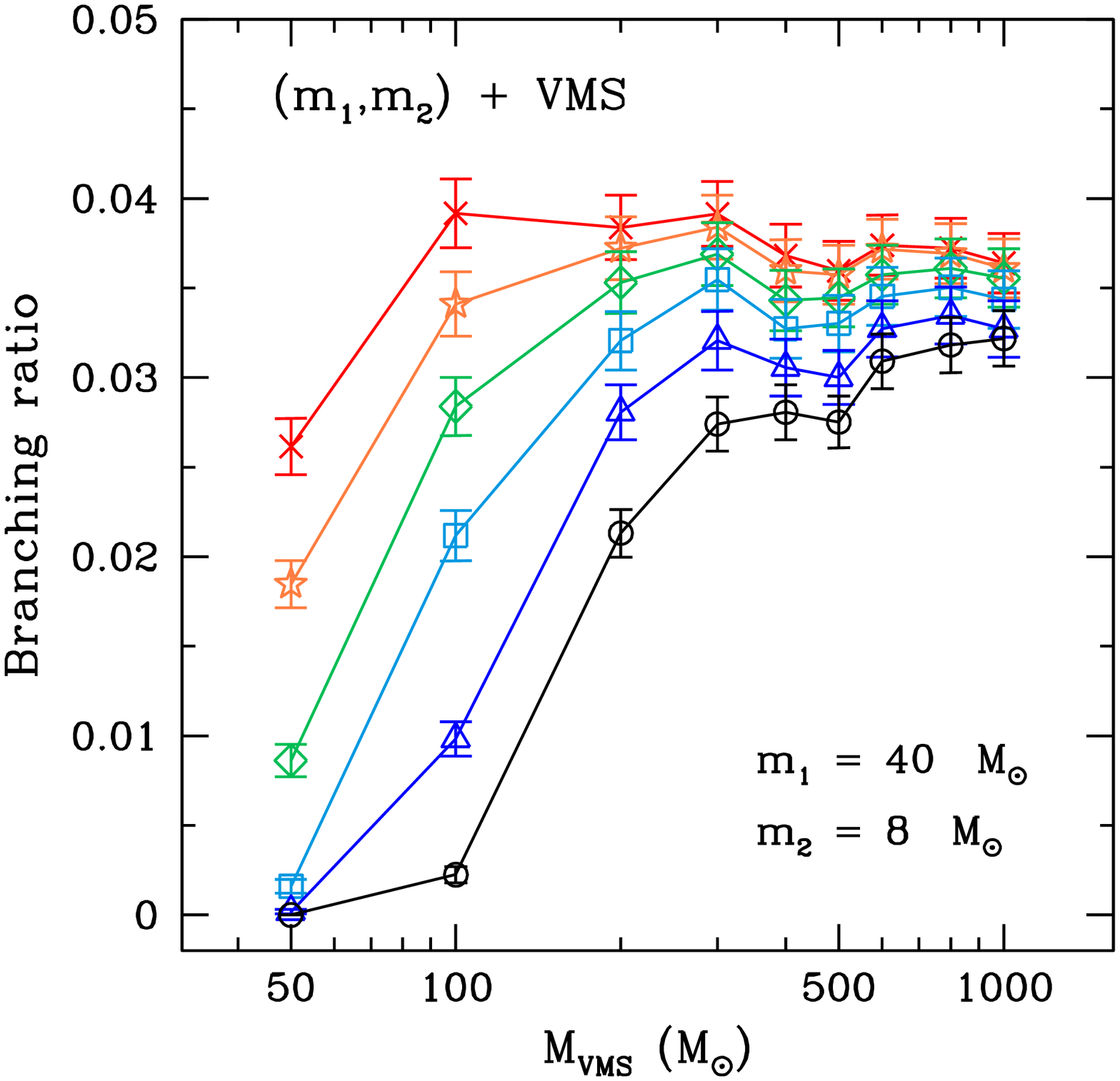}}
\end{minipage}
\caption{{\it Left}: Velocity distributions at infinity for escaping (
  $8 \, \msun$) stars in encounters between a binary consisting of a primary star with mass $m_1 =
  40\, \msun$ and a secondary star with mass $m_2 = 8\, \msun$, and
  a single VMS of mass $M_{\rm VMS}= 50\, \msun , 100\, \msun , 500\, \msun$
  and $1000\, \msun$ (from left to right). The binary semi-major axis is $a$ =
  0.15 \,AU. {\it Right}: The probability of exchange encounters between a $(40,8)\,
\msun$ binary (with $a=0.15$ AU) and a single VMS resulting in ejection
of the $8 \, \msun$ star with a velocity from 200 to $700 \, \kms$ (top to bottom).}
\label{fig:early}
\end{figure}
\begin{figure}[h]
\begin{minipage}[h]{0.44\linewidth}
\center{\includegraphics[width=1\linewidth]{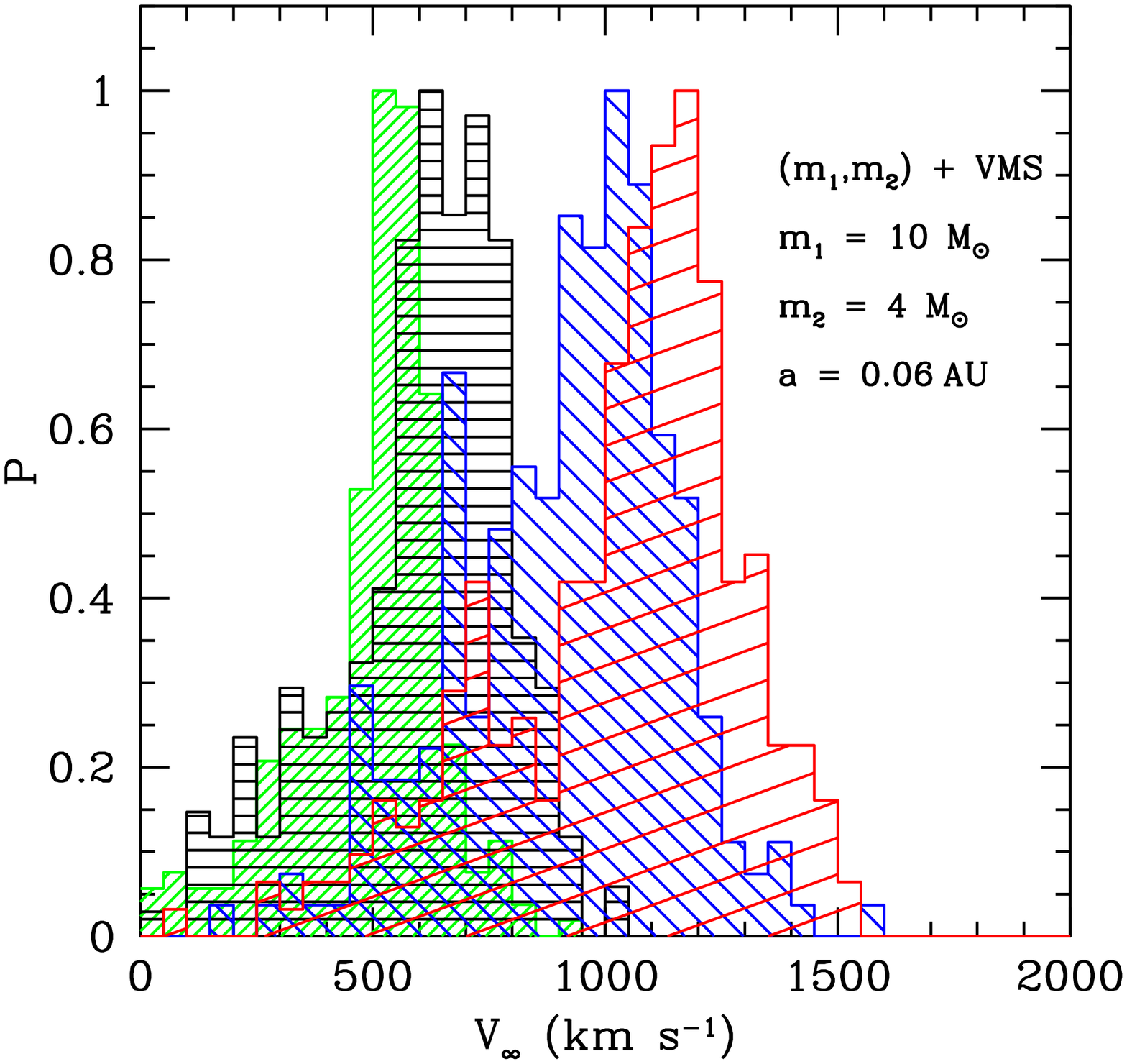}}
\end{minipage}
\hfill
\begin{minipage}[h]{0.44\linewidth}
\center{\includegraphics[width=1\linewidth]{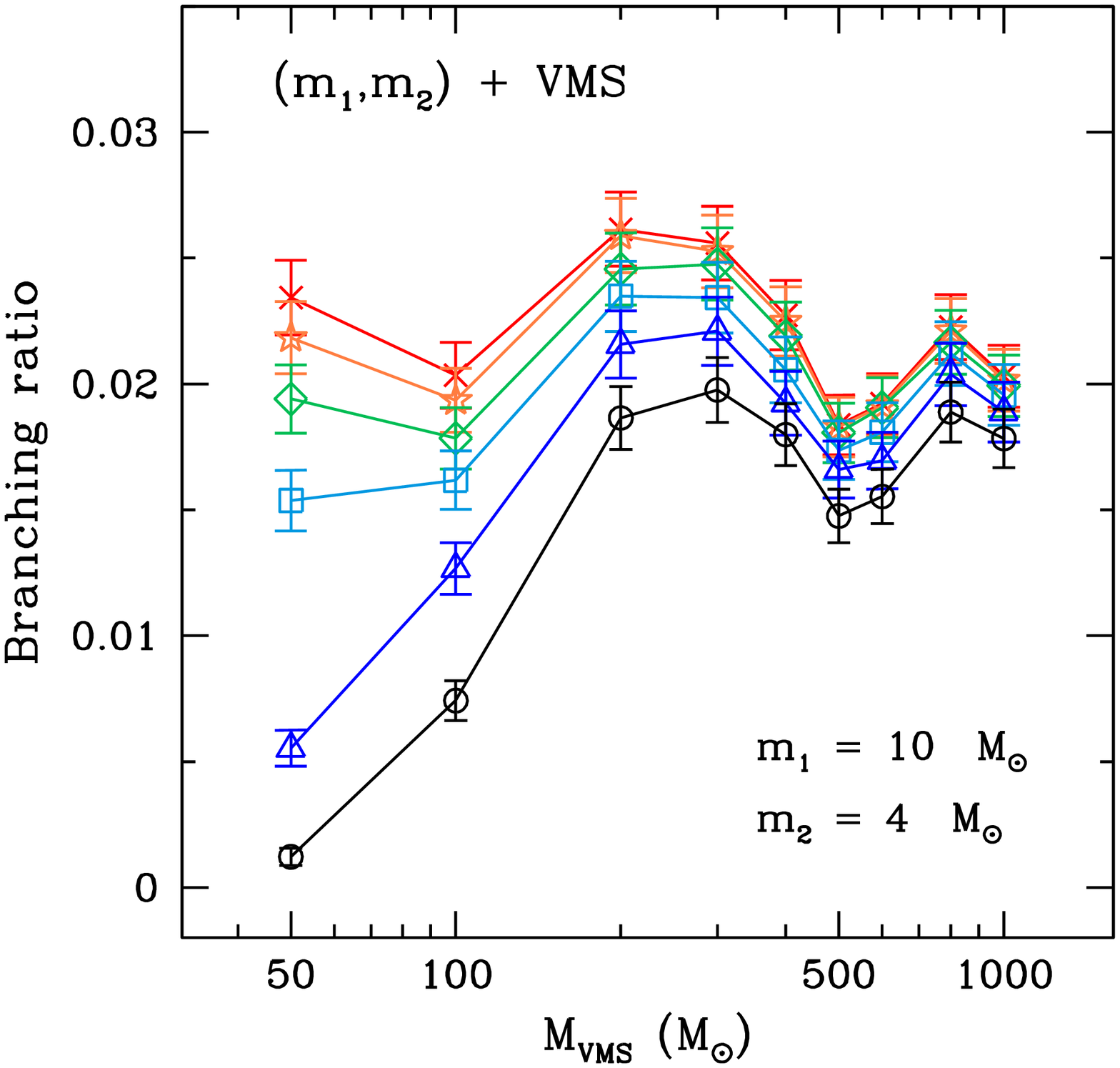}}
\end{minipage}
\caption{{\it Left}: Velocity distributions at infinity for escaping stars in encounters
  between a binary consisting of a primary star with mass $m_1 =
  10\, \msun$ and a secondary star with mass $m_2 = 4\, \msun$, and
  a single VMS of mass $M_{\rm VMS}=50\, \msun$, $M_{\rm VMS}=100\, \msun$,
  $500\, \msun$ and $1000\, \msun$ (from left to right). The binary semi-major
  axis is $a$ = 0.06 \,AU. {\it Right}: The probability of exchange encounters between a $(10,4)\,
  \msun$ binary (with $a=0.06$ AU) and a single VMS resulting in ejection
  of the $4 \, \msun$ star with a velocity from 200 to $700 \, \kms$
  (top to bottom).}
\label{fig:late}
\end{figure}

First, we focus on exchange encounters producing high-velocity early
B-type stars and simulate encounters between a binary consisting of
two main-sequence stars with masses $m_1 = 40 \, \msun$ and $m_2
=8\, \msun$  and a massive compact body, either a VMS or an ordinary
star of mass of $50-100 \, \msun$ (the most massive ordinary stars
formed in clusters with mass of $\simeq 10^3 -10^4 \, \msun$; e.g.
Weidner, Kroupa \& Bonnell \cite{we09}). In order to maximize the
ejection speed, we assume that the binary system is very tight, e.g.
formed by tidal capture. In this case, $a\simeq 3 \, r_1 \simeq
0.15$ AU, where $r_1$ is the radius of the primary star (see
Gvaramadze et al. \cite{gva09}). Fig.\,\ref{fig:early} (left panel)
shows the velocity distribution for $8 \, \msun$ escapers. As
expected, the more massive VMSs are more likely to eject stars with
high velocities. In the right panel of Fig.\,\ref{fig:early} we show
the probability of exchange encounters resulting in ejection of the
$8 \, \msun$ binary component with a velocity from 200 to $700 \,
\kms$ (top to bottom). For $M_{\rm VMS} \geq 100 \, \msun$ about $3$
per cent of all encounters produce runaways with peculiar velocities
$\geq 400 \, \kms$. It can be seen that even an ordinary star of
mass of $50 \, \msun$ can occasionally (in $\sim 1$ per cent of
encounters) produce an escape velocity $\geq 400 \, \kms$. In order
to produce escapers with velocities typical of HVSs ($V_{\infty}
\geq 700 \, \kms$), a VMS of several hundred solar masses is
required. In the case of a $200-300 \, \msun$ VMS, $\geq 2$ per cent
of all encounters result in an escape velocity of $\geq 700 \,
\kms$. This fraction gradually increases to $\geq 3$ per cent for
the more massive VMSs.

Next, we consider exchange encounters producing high-velocity late
B-type stars and simulated encounters between a very tight binary
($m_1 =10 \, \msun$, $m_2 =4 \, \msun$ and $a\simeq 3 r_1 \simeq
0.06$ AU) and a (very) massive star of mass of $50-1000 \, \msun$.
The velocity distributions for $4 \, \msun$ escapers are shown in
Fig.\,\ref{fig:late} (left panel) for four different values of the
VMS mass: $M_{\rm VMS}= 50, 100, 500$ and $1000 \, \msun$. The right
panel of Fig.\,\ref{fig:late} shows that for all values of $M_{\rm
VMS}$ about $2$ per cent of encounters results in peculiar
velocities $\geq 300-400 \, \kms$ and that about the same percentage
of ejected stars attains velocity $\geq 700 \, \kms$ if $M_{\rm VMS}
\geq 200 \, \msun$.

We therefore argue that the origin of high-velocity ($\geq 200-400
\, \kms$) early and late B-type runaway stars and at least some HVSs
could be associated with dynamical encounters between the tightest
massive binaries and VMSs formed in the cores of YMSCs. The future
proper motion measurements for HVSs with GAIA will reveal what
fraction of these extremely high-velocity stars originated in the
Galactic disk.

\section{Acknowledgements}
We are grateful to L.R.Yungelson for useful discussions. VVG
acknowledges the Russian Foundation for Basic Research and the
International Astronomical Union for travel grants and the Deutsche
Forschungsgemeinschaft for partial financial support. AG is
supported by grant NNX07AH15G from NASA. SPZ acknowledges support
from the Netherlands Organization for Scientific Research (NWO under
grant No. 643.200.503) and the Netherlands Research School for
Astronomy (NOVA).

\end{document}